\documentclass[aps,prd,preprint,groupedaddress,preprintnumbers,longbibliography]{revtex4-1}

\usepackage{amsmath}
\usepackage{amssymb}
\usepackage{graphicx}
\usepackage{hyperref}
\usepackage{color}
\newcommand{\p}{\partial}
\newcommand{\hf}{{1\over 2}}
\newcommand{\be}{\begin{equation}}
\newcommand{\br}{\begin{eqnarray}}
\newcommand{\er}{\end{eqnarray}}
\newcommand{\bt}{\begin{tabular}}
\newcommand{\et}{\end{tabular}}
\newcommand{\ee}{\end{equation}}

\newcommand{\dd}{\delta}
\newcommand{\eps}{\epsilon}
\newcommand{\lm}{\Lambda}
\newcommand{\lo}{\Lambda_0}

\newcommand{\etat}{\frac{\eta}{2}}

\newcommand{\nn}{\nonumber}
\newcommand{\ep}{\epsilon} 
\newcommand{\lb}{\left\lbrace}
\newcommand{\rb}{\right\rbrace}
\newcommand{\vev}[1]{\left\langle #1 \right\rangle}
\newcommand{\vvev}[1]{\left\langle\kern-0.3em\left\langle #1
    \right\rangle\kern-0.3em\right\rangle}
\newcommand{\SL}{S_\Lambda}
\newcommand{\WL}{W_\Lambda}
\newcommand{\W}{\mathcal{W}}
\newcommand{\D}{\mathcal{D}}

\begin{document}

\preprint{IMSc/2019/02/01}
\preprint{KOBE-TH-19-02}

\title{Holographic Wilson's RG}


\author{B. Sathiapalan}
\email[]{bala@imsc.res.in}
\affiliation{Institute of
  Mathematical Sciences\\CIT Campus, Tharamani\\ 
  Chennai 600113, India\\and\\Homi Bhabha National Institute\\Training
  School Complex, Anushakti Nagar\\Mumbai 400085, India}
\author{H.~Sonoda}
\email[]{hsonoda@kobe-u.ac.jp}
\affiliation{Physics Department, Kobe
  University\\Kobe 657-8501, Japan}

\date{\today}

\begin{abstract}
  In an earlier paper (arXiv:1706.03371) a holographic form of the
  Exact Renormalization Group (ERG) evolution operator for a
  (perturbed) free scalar field (CFT) in $D$ dimensions was
  formulated.  It was shown to be equivalent, after a change of
  variables, to a free scalar field
  action in $AdS_{D+1}$ spacetime.  We attempt to extend this result
  to a theory where the scalar field has an anomalous dimension.
  Instead of the ERG evolution operator, we examine the generating
  functional with an infrared cutoff, and derive the prescription of
  alternative quantization by using the change of variables introduced
  in the previous paper.  The anomalous dimension is thus related in
  the usual way to the mass of the bulk scalar field.  Computation of
  higher point functions remains difficult in this theory, but should
  be tractable in the large $N$ version. 
 \end{abstract}


\maketitle

\section{Introduction}

The idea of holography has been with us for some time since the
publication of the first papers \cite{tHooft:1993dmi,
  Susskind:1994vu}. It became a mathematically precise idea with the
discovery of the AdS/CFT correspondence
\cite{Maldacena:1997re,Gubser:1998bc,Witten:1998qj,Witten:1998zw}
where an ordinary conformal field theory (N=4 Super Yang Mills) in $D$
flat dimensions is conjectured to be dual to a gravity theory (IIB
Superstrings) in AdS${}_{D+1}$. By now much evidence has been
collected for the correctness of this conjecture. This correspondence
has a natural interpretation in string theory where there is a world
sheet duality that relates open and closed strings. Nevertheless it is
worth exploring to what extent string theory is {\em required} for a
holographic AdS description of a CFT.  String theory may be required
for a UV completion on the gravity side. But the duality itself may be
more general if we are only interested in an effective field theory
description. It is certainly known that in some limits of the
parameter space gravity is sufficient for the correspondence to be
correct.

Another intriguing aspect of this correspondence is the possibility of
interpreting the extra radial dimension as the renormalization scale
of the boundary theory. This gives rise to the idea of ``holographic''
RG, in which radial evolution in the bulk gravity theory is identified
with RG evolution of the boundary theory
\cite{Akhmedov:1998vf,Akhmedov:2010mz,Alvarez:1998wr,Girardello:1998pd,Distler:1998gb,Freedman:1999gp,deBoer:1999tgo,deBoer:1999,Faulkner:2010jy,Klebanov:1999tb,Heemskerk:2010hk,Lizana:2015hqb,Bzowski:2015pba,deHaro:2000vlm}. It
is natural to ask whether this identification can be made more
precise, i.e., whether it is possible to derive the holographic RG
equation from the RG equation of the boundary field theory. In an
earlier paper \cite{Sathiapalan:2017frk} this was answered in the
affirmative for the simple case of a free massless scalar field
theory. It was shown first that the evolution operator for Wilson's
Exact Renormalization Group
(ERG)\cite{Wilson:1973jj,Wegner:1972ih,Wilson:1993dy}, in the simpler
Polchinski form \cite{Polchinski:1983gv}, could be written as a
functional integral of a $D+1$ dimensional field theory. (See
\cite{Becchi:1996an,Bagnuls:2001pr,Igarashi:2009tj,Rosten:2010vm} for
reviews on ERG.)  A change of field variables then transformed this
operator into the action for a free scalar field theory in
$AdS_{D+1}$. A contact was thus made with the standard AdS/CFT methods
for the calculation of two-point correlators. One important point is
that the bulk field took the value of the boundary field at the
boundary (rather than the source) so this is more naturally understood
as the {\em alternative quantization} procedure introduced in
\cite{Klebanov:1999tb}.

In the present paper we change the course of approach a little by
considering the generating functional of correlation functions with an
infrared cutoff \cite{Morris:1993qb,Morris:1994ie,Igarashi:2016qdr}
instead of the ERG evolution operator.  The generating functional is
closely related to a Wilson action, and it reduces to the ordinary
generating functional in the limit of the vanishing infrared cutoff.
We follow section 5.2 of \cite{Sathiapalan:2017frk} by introducing an
elementary scalar field of scale dimension between $(D-2)/2$ and $D/2$
to represent a composite field.  We then construct a {\em quadratic}
Wilson action that gives the expected two-point function with an
anomalous dimension. Normally anomalous dimensions arise due to
interactions.  However it is very hard to write down fixed point
Wilson actions with interactions and anomalous dimension. Therefore to
clarify the role of the anomalous dimension in the map from an ERG
equation to an AdS evolution equation we consider a simpler model of a
Gaussian theory with anomalous dimension.  It solves the standard
fixed point ERG equation with anomalous dimension. The mapping
techniques used for this simple model should be applicable in the more
realistic case of an interacting fixed point also.

Having constructed a fixed point Wilson action, we construct a
corresponding generating functional $W_\Lambda [J]$ with an infrared
cutoff $\Lambda$ following a recipe well known in the ERG literature.
In the infrared limit $\Lambda \to 0+$, $W_\Lambda [J]$ becomes the
generating functional of the connected correlation functions.  Since
we are ignoring interactions, we obtain
\[
  \lim_{\Lambda \to 0+} W_\Lambda [J] = \frac{1}{2} \int_{p,q} J(-p)
    \vev{\phi (p) \phi (q)} J(-q)\,,
\]
where the two-point function
\[
  \vev{\phi (p) \phi (q)} = \delta (p+q) \cdot \frac{1}{p^{2-\eta}}
\]
has the anomalous dimension $\eta$ of the scalar field.

Let us sketch briefly how the AdS space arises from ERG without going
much into technical details.  According to ERG, the cutoff dependence
of the generating functional is given by a diffusion equation (in the
main text introduced as (\ref{ERGdiffeqW}) or more precisely,
including an anomalous dimension parameter $\eta$, as
(\ref{ERGdiffeqWtilde})).  The equation is solved by the Gaussian
integral formula
\[
  e^{W_{\Lambda_2} [J]}
  = \int [dJ'] \exp \left[ W_{\Lambda_1} [J+J'] - \frac{1}{2} \int_p
    \frac{J' (p) J' (-p)}{R_{\Lambda_1} (p) - R_{\Lambda_2} (p)} 
  \right]\,,
\]
where $\Lambda_2 < \Lambda_1$, and $R_\Lambda (p)$ is an IR cutoff
function.  $W_\Lambda [J]$ is quadratic in $J$, and we can write the
above in the form
\[
  e^{W_{\Lambda_2} [J]}
  = \int [d\varphi] \exp \left[ - \frac{1}{2} \int_p \frac{\varphi (p) \varphi
      (-p)}{G_{1/\Lambda_2} (p) - G_{1/\Lambda_1} (p)} + \int_p
    \varphi (p)
    J(-p) + \cdots
  \right]\,,
\]
where $\varphi$ is a rescaled $J'$, and we have suppressed a term
quadratic in $J$.  $G_{1/\Lambda} (p)$ is the two-point function with
an IR cutoff $\Lambda$.  We can write the quadratic term as a
functional integral over the field $y(z,p)$ that interpolates $J' (p)$
at $z = \frac{1}{\Lambda_1}$ and $0$ at $z=\frac{1}{\Lambda_2}$:
\[
  \exp \left[ - \frac{1}{2} \int_p \frac{\varphi (p) \varphi
      (-p)}{G_{1/\Lambda_2} (p) - G_{1/\Lambda_1} (p)} \right]
  = \int \mathcal{D} y\, \exp \left[ - \frac{1}{2}
    \int_{\frac{1}{\Lambda_1}}^{\frac{1}{\Lambda_2}} dz
    \int_p \frac{\partial_z y(z,p) \partial_z y(z,-p)}{\partial_z G_z
      (p)} \right]\,.
\]
$z$ gives the radial coordinate of AdS${}_{D+1}$:
$z = \frac{1}{\Lambda_1} = \ep$ is the radius of the boundary, and
$z = \frac{1}{\Lambda_2} = z_0$ is to be taken to infinity.  Using the
same change of field variables
\[
  y(z,p) \longrightarrow Y(z,p)
\]
that we introduced in 2.3 of
\cite{Sathiapalan:2017frk} (given precisely by (\ref{y-to-Y})),
we can rewrite the generating functional in the AdS form:
\[
  \int \mathcal{D} Y\, \exp \left( S_{\mathrm{AdS}} [Y] \right)\,,
\]
where $S_{\mathrm{AdS}} [Y]$, given by (\ref{AdSaction}), is the
action of a massive free field in the space AdS${}_{D+1}$ with the
metric
\[
  \frac{dz^2 + d\vec{x} \cdot d\vec{x}}{z^2}\,.
\]
We still need to integrate over the field $\varphi (p) \sim Y(\ep,p)$ at
the boundary; we thus reproduce the prescription of the alternative
quantization \cite{Klebanov:1999tb} (reviewed nicely in Appendix of
\cite{Faulkner:2010jy}) for computing the two-point function.

Unlike the usual Feynman diagram approach where one integrates over
all momenta in a loop, in the Wilsonian approach only modes above a
scale $\lm$ are integrated out. In the holographic description, where
the radial coordinate is a measure of the scale, this would correspond
to integrating out fields beyond a certain value of the radius. This
is precisely the notion emphasized by holographic RG (see for example
\cite{Faulkner:2010jy}) where the picture is of the boundary moving
inward as one proceeds toward the IR.  This is the underlying reason
why ERG techniques are able to reproduce holographic results.

This paper is organized as follows. In Sec. \ref{section:background}
we overview the ERG formalism to introduce a quadratic Wilson action
that gives a two-point function with an anomalous dimension.  We then
introduce a corresponding generating functional with an infrared
cutoff in Sec. \ref{section:derivation} to apply the change of
variables of \cite{Sathiapalan:2017frk}.  By a judicious adjustment of
the change of variables, we can derive the prescription of alternative
quantization. We discuss our
result and method in Sec. \ref{section:discussion}. In Sec. \ref{elucidation} we give some background
regarding anomalous dimension in ERG and its connection with the
change of variables used in \cite{Sathiapalan:2017frk}.
Sec. \ref{nonlinear} contains a preliminary discussion of a situation
where one might obtain a nontrivial (i.e. cubic and higher order )
bulk action starting from a generalized ERG equation.   We conclude
the paper in Sec. \ref{section:conclusion}.

\section{Background}\label{section:background}

\subsection{ERG formalism}

For the convenience of the reader, we would like to collect relevant
background material from the exact renormalization group formalism (ERG).

Let $\SL [\phi]$ be a Wilson action of a generic scalar field theory.
To preserve physics independent of $\Lambda$, we impose the ERG
differential equation
\begin{equation}
- \Lambda \frac{\partial}{\partial \Lambda} e^{\SL [\phi]} = \int_p
\left[ \frac{\Delta_\Lambda (p)}{K_\Lambda (p)} \phi (p)
  \frac{\delta}{\delta \phi (p)}
 + \frac{\Delta_\Lambda (p)}{p^2} \frac{1}{2} \frac{\delta^2}{\delta
   \phi (p) \delta \phi (-p)} \right]\, e^{\SL [\phi]}\,,\label{ERGdiffeq}
\end{equation}
where
\begin{equation}
\Delta_\Lambda (p) \equiv \Lambda \frac{\partial}{\partial \Lambda}
K_\Lambda (p)\,.
\end{equation}
The cutoff function $K_\Lambda (p)$ has three properties:
\begin{enumerate}
\item $K_\Lambda (0) = 1$,
\item it is of order $1$ for $p^2 < \Lambda^2$,
\item it approaches $0$ rapidly for $p^2 \gg \Lambda^2$.
\end{enumerate}
For example, we can take
$K_\Lambda (p) = K(p/\Lambda) = \exp \left(- p^2/\Lambda^2\right)$ as
shown in Fig.~1.
($k_\Lambda (p) = K(p/\Lambda)\left(1-K(p/\Lambda)\right)$ and
$R_\Lambda (p) = \Lambda^2 R(p/\Lambda) = p^2 K(p/\Lambda)/\left(1 -
  K(p/\Lambda)\right)$ are respectively defined by
Eqs.~(\ref{kLambda}) and (\ref{RLambda}) below.)
\begin{figure}[htb]
\centering
\includegraphics[width=8cm]{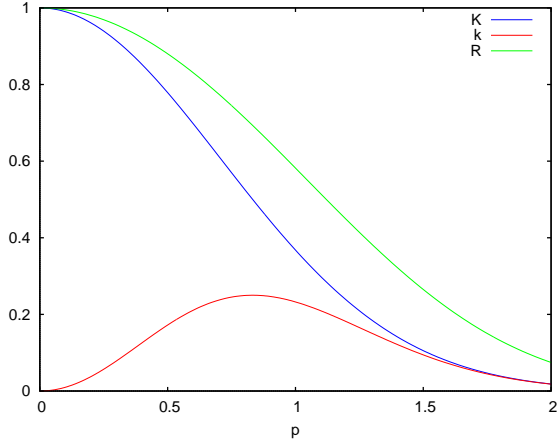}
\caption{We plot $K (p) = e^{-p^2},\, k (p) = e^{-p^2} (1 - e^{-p^2})$,
  and $R (p) = p^2 \frac{1}{e^{p^2}-1}$}
\end{figure}

We denote the correlation functions by
\begin{equation}
\vev{\phi (p_1) \cdots \phi (p_n)}_{\SL} \equiv \int [d\phi] \phi
(p_1) \cdots \phi (p_n)\, e^{\SL [\phi]}\,.
\end{equation}
The ERG differential equation (\ref{ERGdiffeq}) implies that
the correlation functions defined below are independent of $\Lambda$:
\begin{equation}
\vvev{\phi (p_1) \cdots \phi (p_n)} 
 \equiv \prod_{i=1}^n
\frac{1}{K_\Lambda (p_i)}\cdot \vev{\exp \left( - \frac{1}{2} \int_p
    \frac{k_\Lambda (p)}{p^2}
    \frac{\delta^2}{\delta \phi (p) \delta \phi (-p)} \right)\phi
  (p_1) \cdots \phi (p_n)}_{\SL}\,,\label{double-bracket}
\end{equation}
where
\begin{equation}
  k_\Lambda (p) \equiv K_\Lambda (p) \left( 1 - K_\Lambda (p)\right)\,.
  \label{kLambda}
\end{equation}
The modification of the two-point function by $k_\Lambda (p)/p^2$ does
not affect the physics in the infrared as long as $k_\Lambda (p)$
vanishes as $p^2$ at $p^2=0$.  (In other words, $k_\Lambda (p)/p^2$
does not correspond to the propagation of a free massless particle.)
We explain a little more on the correlation functions in double
brackets in Appendix \ref{appendix-double}.

We then define the generating functional $\mathcal{W} [\mathcal{J}]$ of the
connected correlation functions by
\begin{equation}
  e^{\mathcal{W} [\mathcal{J}]} \equiv \sum_{n=0}^\infty \frac{1}{n!} \int_{p_1,
    \cdots, p_n} \mathcal{J} ( - p_1) \cdots \mathcal{J} (- p_n)
  \vvev{\phi (p_1) \cdots \phi (p_n)}\,.
\end{equation}
This can be written as a functional integral in the presence of a
source term:
\begin{align}
  e^{\mathcal{W} [\mathcal{J}]} &= \sum_{n=0}^\infty \frac{1}{n!}
  \int_{p_1,\cdots,p_n} \prod_{i=1}^n \frac{\mathcal{J}(
    -p_i)}{K_\Lambda (p_i)}\cdot \vev{\exp \left( - \int_p
      \frac{k_\Lambda (p)}{p^2} \frac{1}{2} \frac{\delta^2}{\delta
        \phi (p) \delta \phi
        (-p)} \right) \phi (p_1) \cdots \phi (p_n)}_{\SL} \nn\\
  &= \vev{\exp \left( - \int_p \frac{k_\Lambda (p)}{p^2} \frac{1}{2}
      \frac{\delta^2}{\delta \phi (p) \delta \phi (-p)} \right) \exp
    \left( \int_p
      \frac{\mathcal{J}(-p)}{K_\Lambda (p)} \phi (p) \right) }_{\SL}\nn\\
  &= \vev{\exp \left( \int_p
      \left(\frac{\mathcal{J}(-p)}{K_\Lambda (p)} \phi (p) -
        \frac{1}{2} \frac{1}{R_\Lambda (p)} \mathcal{J}(p) \mathcal{J}
        (-p)\right)\right) }_{\SL}\,,
\end{align}
where
\begin{equation}
R_\Lambda (p) \equiv \frac{p^2 K_\Lambda (p)^2}{k_\Lambda (p)} =
\frac{p^2 K_\Lambda (p)}{1 - K_\Lambda (p)}\,.\label{RLambda}
\end{equation}

We now define a field
\begin{equation}
J (p) \equiv \phi (p) \frac{R_\Lambda (p)}{K_\Lambda (p)} = \phi
(p) \frac{p^2}{1 - K_\Lambda (p)}\label{Jprime}
\end{equation}
and following \cite{Morris:1993qb,Morris:1994ie} introduce
\begin{equation}
\WL [J] \equiv \SL [\phi] + \frac{1}{2} \int_p \frac{J (p) J
  (-p)}{R_\Lambda (p)}\,.\label{WL-def}
\end{equation}
We then obtain
\begin{align}
e^{\W [\mathcal{J}]} &= \int [dJ] \exp \left( \WL [J] - \frac{1}{2} \int_p
  \frac{1}{R_\Lambda (p)} \left( J (p) - \mathcal{J}(p) \right)\left( J (-p) -
    \mathcal{J} (-p) \right)\right)\nn\\
&= \int [dJ] \exp \left( \WL [J+\mathcal{J}] - \frac{1}{2} \int_p
  \frac{1}{R_\Lambda (p)} J (p) J (-p)\right)\,.\label{WfromWL}
\end{align}
Since
\begin{equation}
\lim_{\Lambda \to 0+} K_\Lambda (p) = 0\,,
\end{equation}
we obtain
\begin{equation}
\lim_{\Lambda \to 0+} R_\Lambda (p) = 0\,.\label{Rzero}
\end{equation}
Hence, we obtain \cite{Morris:1993qb,Morris:1994ie,Sonoda:2017rro}
\begin{equation}
\W [\mathcal{J}] = \lim_{\Lambda \to 0+} \WL [\mathcal{J}]\,.\label{WL-limit}
\end{equation}

We can think of $\WL [J]$ as the generating functional with an IR
cutoff $\Lambda$.  Its
$\Lambda$-dependence can be obtained from the ERG differential
equation (\ref{ERGdiffeq}) as
\begin{equation}
- \Lambda \frac{\partial}{\partial \Lambda} e^{\WL [J]} = \int_p
\Lambda \frac{\partial R_\Lambda (p)}{\partial \Lambda} \frac{1}{2}
\frac{\delta^2}{\delta J (p) \delta J (-p)} e^{\WL [J]}\,,\label{ERGdiffeqW}
\end{equation}
which can be solved as
\begin{equation}
e^{W_{\Lambda_2} [J]} = \int [dJ'] \exp \left[ W_{\Lambda_1} [J' + J]
 - \frac{1}{2} \int_p \frac{1}{R_{\Lambda_1} (p) - R_{\Lambda_2}
  (p) }J' (p) J' (-p)  \right]\,.\label{WL-dep}
\end{equation}
Since $R_{\Lambda_1} (p) - R_{\Lambda_2} (p)$ is non-vanishing mainly
for $\Lambda_2 < p < \Lambda_1$, the above equation implies that we
obtain $W_{\Lambda_2} [J]$ from $W_{\Lambda_1} [J]$ by integrating
fluctuations of momenta between $\Lambda_2$ and $\Lambda_1$.  In the
limit $\Lambda \to 0+$, all the momentum modes are integrated to give
(\ref{WL-limit}).

\subsection{Quadratic Wilson action with an anomalous dimension}

We now consider a simple quadratic Wilson action with an anomalous
dimension:
\begin{equation}
  \SL [\phi] = - \frac{1}{2} \int_p \frac{p^2}{K_\Lambda (p)}
  \frac{1}{1 + K_\Lambda (p) \left( \left(\frac{p}{\mu}\right)^\eta - 1 \right)}
  \phi (p) \phi (-p)\,,\label{quad-action}
\end{equation}
where $\eta = 2 \gamma\, (0 < \gamma < 1)$ is a positive anomalous
dimension, and $\mu$ is an arbitrary reference momentum scale.  This
action reproduces a two-point function with an anomalous dimension:
\begin{align}
\vvev{\phi (p) \phi (q)} &= \frac{1}{K_\Lambda (p)^2} \left( \vev{\phi
                           (p) \phi (q)}_{\SL} -
                           \frac{k_\Lambda (p)}{p^2} \delta (p+q) \right)\notag\\
&= \frac{1}{p^2 \left(\frac{p}{\mu}\right)^{-\eta}} \, \delta (p+q)\,.
\end{align}

The corresponding generating functional with an IR cutoff $\Lambda$ is given by
\begin{align}
\WL [J] &\equiv \SL [\phi] + \frac{1}{2} \int_p \frac{J (p)
  J(-p)}{R_\Lambda (p)} \nn\\
&=  \frac{1}{2} \int_p J (p) J (-p) \frac{1}{p^2
  \left(\frac{p}{\mu}\right)^{-\eta} + R_\Lambda (p)}\,,\label{WL-anom}
\end{align}
where
$1/\left( p^2 \left(p/\mu\right)^{-\eta} + R_\Lambda (p)\right)$
is the high-momentum propagator, or the two-point function with an infrared
cutoff.  Using (\ref{Rzero}), we get
\begin{equation}
\W [J] = \lim_{\Lambda \to 0} \WL [J] = \frac{1}{2} \int_p J(p)
\frac{1}{p^2 \left(\frac{p}{\mu}\right)^{-\eta}} J (-p)\,.
\end{equation}

Please observe that for small $p \ll \Lambda$ the action is
approximately given by
\begin{equation}
\SL [\phi] \simeq - \frac{1}{2} \int_p
\frac{p^2}{\left(\frac{p}{\mu}\right)^\eta}\, \phi (p) \phi (-p)\,,
\end{equation}
which is a non-analytic (non-local) action.  We have two comments:
\begin{enumerate}
\item For an elementary field $\phi$, we expect the action is analytic
  at zero momentum, and any non-analyticity comes from interactions.
  As $\phi$, we have composite fields in mind.
\item For example, in the massless free theory, the composite field
  $\frac{1}{2} \phi^2$ has scale dimension $D-2$ so that the anomalous
  dimension, compared with the free elementary field, is
\[
\gamma = \frac{D-2}{2} \Longleftrightarrow
D-2 = \frac{D-2}{2} + \gamma\,.
\]
$0 < \gamma < 1$ implies $2 < D < 4$.  This is an example of our
$\phi$.
\end{enumerate}
We will discuss the first point further in Sec. \ref{section:discussion}.

\subsection{ERG formalism with explicit dependence on the anomalous dimension}

One drawback of our choice $W_\Lambda$ (\ref{WL-anom}) is that the ERG
equation (\ref{ERGdiffeqW}) it satisfies shows no sign of the
anomalous dimension $\eta$ contained in $W_\Lambda$.  In the usual
AdS/CFT calculations the anomalous dimension $\eta$ is introduced as a
mass term in the AdS equations.  Similarly, we would like to introduce
$\eta$ explicitly in the ERG equation.  (More will be discussed later
in Sec. \ref{section:identification}.)  

Given $W_\Lambda [J]$ satisfying (\ref{ERGdiffeqW}), let us define
\begin{equation}
\tilde{W}_\Lambda [J] \equiv W_\Lambda \left[
  \left(\frac{\Lambda}{\mu}\right)^{- \frac{\eta}{2}} J \right]\,.
\end{equation}
Then, (\ref{ERGdiffeqW}) implies that $\tilde{W}_\Lambda [J]$
satisfies the alternate ERG equation with an explicit dependence on
$\eta$:
\begin{equation}
- \Lambda \frac{\partial}{\partial \Lambda} e^{\tilde{W}_\Lambda [J]}
= \int_p \left[ \frac{\eta}{2} J(p) \frac{\delta}{\delta J(p)} +
  \left( \Lambda \frac{\partial \tilde{R}_\Lambda (p)}{\partial \Lambda} -
    \eta \tilde{R}_\Lambda (p)\right) \frac{1}{2} \frac{\delta^2}{\delta J(p)
    \delta J(-p)} \right] e^{\tilde{W}_\Lambda [J]}\,,\label{ERGdiffeqWtilde}
\end{equation}
where
\begin{equation}
\tilde{R}_\Lambda (p) \equiv \left(\frac{\Lambda}{\mu}\right)^\eta
R_\Lambda (p)\,.
\end{equation}
To derive (\ref{ERGdiffeqWtilde}) from (\ref{ERGdiffeqW}), we have
used
\begin{equation}
\Lambda \frac{\partial R_\Lambda (p)}{\partial \Lambda} =
\left(\frac{\Lambda}{\mu}\right)^{-\eta} \left( \Lambda \frac{\partial
    \tilde{R}_\Lambda (p)}{\partial \Lambda} - \eta \tilde{R}_\Lambda
  (p) \right)\,.
\end{equation}

The integral formula (\ref{WL-dep}) implies the corresponding integral
formula:
\begin{align}
&e^{\tilde{W}_\Lambda \left[
    \left(\frac{\Lambda}{\mu}\right)^{\frac{\eta}{2}} J \right]}\nn\\
&= \int [dJ'] \exp \left[ W_{\Lambda'} \left[
    \left(\frac{\Lambda'}{\mu}\right)^{\frac{\eta}{2}} (J'+J)
  \right]
 - \frac{1}{2} \int_p \frac{ J' (p) J' (-p)}{ (\mu/\Lambda')^\eta
  \tilde{R}_{\Lambda'} (p) - (\mu/\Lambda)^\eta \tilde{R}_\Lambda (p)}\right]\,.
\label{WtildeL-integral}
\end{align}
(\ref{Rzero}) can now be written as
\begin{equation}
\lim_{\Lambda \to 0+} \left(\frac{\Lambda}{\mu}\right)^{-\eta}
\tilde{R}_\Lambda (p) = 0\,,\label{Rtildezero}
\end{equation}
and we obtain, from (\ref{WL-limit}),
\begin{equation}
\mathcal{W} [\mathcal{J}] = \lim_{\Lambda \to 0+} \tilde{W}_\Lambda
\left[ \left(\frac{\Lambda}{\mu}\right)^{\frac{\eta}{2}} \mathcal{J}
\right]\,.\label{WtildeL-limit}
\end{equation}

So far, we have not assumed that $W_\Lambda [J]$ is quadratic and
given by (\ref{WL-anom}).  Let us now assume it so that
\begin{equation}
  \tilde{W}_\Lambda [J] = \frac{1}{2} \int_p \frac{J(p) J(-p)}{p^2
    \left(\frac{\Lambda}{p}\right)^\eta + \tilde{R}_\Lambda
    (p)}\,.\label{WtildeL-quadratic} 
\end{equation}
This has no more dependence on the reference momentum $\mu$.  
For the quadratic $\tilde{W}_\Lambda [J]$, (\ref{ERGdiffeqWtilde})
reduces to
\begin{equation}
- \Lambda \frac{\partial \tilde{W}_\Lambda [J]}{\partial \Lambda} =
\int_p \left[ \frac{\eta}{2} J(p) \frac{\delta \tilde{W}_\Lambda
    [J]}{\delta J(p)} + \left(\Lambda \frac{\partial \tilde{R}_\Lambda
      (p)}{\partial \Lambda} - \eta \tilde{R}_\Lambda (p) \right)
  \frac{1}{2} \frac{\delta \tilde{W}_\Lambda [J]}{\delta J(p)}
\frac{\delta \tilde{W}_\Lambda [J]}{\delta J(-p)} \right]\,.
\end{equation}
(\ref{WtildeL-limit}) gives
\begin{equation}
\mathcal{W} [J] = \lim_{\Lambda \to 0+} \tilde{W}_\Lambda \left[
  \left(\frac{\Lambda}{\mu}\right)^{\frac{\eta}{2}} J\right] =
\frac{1}{2} \int_p J(p) \,\frac{1}{p^2
  \left(\frac{p}{\mu}\right)^{-\eta}} \,J(-p)\,.
\end{equation}

As the original ERG equation (\ref{ERGdiffeqW}) does not single out a
particular $\eta$ in (\ref{WL-anom}), the ERG equation
(\ref{ERGdiffeqWtilde}) with an explicit dependence on $\eta$ has a
more general quadratic solution given by
\begin{equation}
\tilde{W}_\Lambda [J] = \frac{1}{2} \int_p \frac{J(p) J(-p)}{p^2
  \left(\frac{\Lambda}{p}\right)^\eta
  \left(\frac{\mu}{p}\right)^{\Delta \eta} + \tilde{R}_\Lambda (p)}\,,
\end{equation}
where $\Delta \eta$ is an arbitrary shift of the anomalous dimension.
Given (\ref{ERGdiffeqWtilde}), what makes (\ref{WtildeL-quadratic})
stand out?  It is the relation to an RG fixed point.  Let us elaborate
on this a little.  We assume that the cutoff function
$\tilde{R}_\Lambda (p)$ has a particular cutoff dependence:
\begin{equation}
\tilde{R}_\Lambda (p) = p^2 r (p/\Lambda)\,.\label{RLambda-dep}
\end{equation}
We define a dimensionless field
\begin{equation}
\bar{J} (\bar{p}) \equiv \Lambda^{\frac{D-2}{2}} J (\bar{p} \Lambda)
\end{equation}
with the dimensionless momentum $\bar{p}$.  We then define
\begin{equation}
\bar{W} [\bar{J}] \equiv \tilde{W}_\Lambda [J]
= \frac{1}{2} \int_{\bar{p}} \frac{\bar{J} (\bar{p}) \bar{J}
  (-\bar{p})}{\bar{p}^{2 - \eta} + R (\bar{p})}\,,
\label{Wbar}
\end{equation}
where
\begin{equation}
R(\bar{p}) \equiv \bar{p}^2 r (\bar{p})\,.
\end{equation}
We find that $\bar{W} [\bar{J}]$ is a fixed point action, satisfying
the equation for scale invariance
\begin{align}
  &\int_{\bar{p}} \left[ \bar{J} (- \bar{p}) \left( - \bar{p}
      \cdot \partial_{\bar{p}} - \frac{D+2}{2} + \frac{\eta}{2}
    \right) \frac{\delta}{\delta
      \bar{J}(-\bar{p})} e^{\bar{W} [\bar{J}]}\right.\nn\\
  &\quad\left. + \frac{1}{2} \left(- \bar{p} \cdot \partial_p + 2 -
      \eta \right) R 
    (\bar{p}) \cdot \frac{\delta^2}{\delta \bar{J}
      (\bar{p}) \delta \bar{J} (- \bar{p})} e^{\bar{W}
    [\bar{J}]} \right] = 0\,.\label{Wbar-scale}
\end{align}
At the same time it also satisfies the equation for special conformal
invariance \cite{Sonoda:2017zgl}
\begin{align}
  &\int_p \bar{J} (-p) \left( - \bar{p}_\nu \frac{\partial^2}{\partial
      \bar{p}_\mu \partial \bar{p}_\nu} + \frac{1}{2} \bar{p}_\mu
    \frac{\partial^2}{\partial \bar{p}_\nu \partial \bar{p}_\nu} + \left(-
      \frac{D+2}{2} + \frac{\eta}{2}\right) \frac{\partial}{\partial
      \bar{p}_\mu} \right)
  \frac{\delta}{\delta \bar{J} (-\bar{p})} e^{\bar{W}[\bar{J}]}\nn\\
  & + \frac{1}{2} \int_{\bar{p}}  \left(- \bar{p} \cdot \partial_p + 2
    - \eta \right) R (\bar{p}) \cdot \frac{\partial}{\partial
    \bar{p}_\mu} \left(\frac{\delta^2}{\delta \bar{J} (-\bar{p}) \delta
    \bar{J} (\bar{q})} e^{\bar{W} [\bar{J}]}
\right)\Big|_{\bar{q}=\bar{p}} = 0\,.
\label{Wbar-conf}
\end{align}
In Appendix \ref{appendix-Wbar} we show how to derive (\ref{Wbar})
either from (\ref{Wbar-scale}) or from (\ref{Wbar-conf}).

\section{Derivation of the alternative quantization}\label{section:derivation}

To simplify our notation, we omit the tilde from $\tilde{W}$
altogether, and consider
\begin{equation}
W_\Lambda [J] = \frac{1}{2} \int_p \frac{J(p) J(-p)}{p^2
  \left(\frac{\Lambda}{p}\right)^\eta + R_\Lambda (p)}\,.\label{WL-new}
\end{equation}
Let $\Lambda_1 = \frac{1}{\ep}$ be a large cutoff, and $\Lambda_2 =
\frac{1}{z_0}$ be a small cutoff, compared with a reference momentum
scale $\mu$.  Our goal is to derive the prescription of the
alternative quantization of AdS/CFT from the integral formula
(\ref{WtildeL-integral}):
\begin{align}
\exp \left[ W_{1/z_0} \left[ (\mu z_0)^{-\frac{\eta}{2}} J
  \right]\right] &= \int [dJ'] \exp \left[ W_{1/\ep} \left[(\mu
    \ep)^{-\frac{\eta}{2}} (J'+J)\right]\right.\nn\\
&\quad\left. - 
  \frac{1}{2} \int_p \frac{1}{(\mu \ep)^\eta R_{1/\ep} (p) - (\mu
    z_0)^\eta R_{1/z_0} (p)} J' (p) J'
  (-p) \right]\,.
\end{align}
Since (\ref{WL-new}) is quadratic, we can expand and rewrite the above as
\begin{align}
&\exp \left[ W_{1/z_0} \left[ (\mu z_0)^{-\frac{\eta}{2}} J
  \right]\right]  \nn\\
&= \int [dJ'] \exp \left[ - \frac{1}{2} \int_p J'
                    (p) J' (-p) \lb \frac{1}{(\mu \ep)^\eta R_{1/\ep}
                      (p) - (\mu z_0)^\eta R_{1/z_0} (p)} -
                    \frac{1}{p^2 \left(\frac{p}{\mu}\right)^{-\eta} +
                      (\mu \ep)^\eta R_{1/\ep} (p)} \rb\right.\nn\\
&\qquad \left. + \int_p J (p) \frac{J' (-p)}{p^2
                    \left(\frac{p}{\mu}\right)^{-\eta} + (\mu
                    \ep)^\eta R_{1/\ep}
                    (p)} + \frac{1}{2} \int_p \frac{J(p) J(-p)}{p^2
                    \left(\frac{p}{\mu}\right)^{-\eta} + (\mu
                    \ep)^\eta R_{1/\ep} (p)} \right]\,.
\end{align}
Rewriting the integration variable as
\[
\varphi (p) \equiv \frac{J' (p)}{ p^2 (p/\mu)^{-\eta} + (\mu \ep)^\eta R_{1/\ep}
  (p)}
\]
to normalize the $J\varphi$ term, we obtain
\begin{align}
\exp \left[ W_{1/z_0} \left[ (\mu z_0)^{-
      \frac{\eta}{2}} J \right]\right]  &= \int [d\varphi] \exp \left[ -
  \frac{1}{2} \int_p \frac{\varphi (p) \varphi (-p)}{G_{z_0} (p) - G_\ep (p)}\right.\nn\\
&\quad \left. + \int_p J(p) \varphi (-p) + \frac{1}{2} \int_p J(p) J(-p)
  G_\ep (p) \right]\,,\label{W-start}
\end{align}
where we have defined
\begin{equation}
G_z (p) \equiv \frac{1}{p^2 \left(\frac{p}{\mu}\right)^{-\eta} + (\mu
  z)^\eta R_{1/z} (p)} \,.
\end{equation}
Note that the parameter
\begin{equation}
  z \equiv \frac{1}{\Lambda}
\end{equation}
is the inverse cutoff, and it will be interpreted as a coordinate of
the AdS${}_{D+1}$ space.  $z$ takes a value between
$\frac{1}{\Lambda_1} = \ep$(small) and
$\frac{1}{\Lambda_2} = z_0$(large). With $G_z (p)$, we can write
\begin{equation}
W_{1/z} \left[ (\mu z)^{- \frac{\eta}{2}} J \right] = \frac{1}{2}
\int_p G_z (p) J(p) J(-p)\,.
\end{equation}
Using (\ref{Rtildezero})
\begin{equation}
\lim_{z \to +\infty} (\mu z)^\eta R_{1/z} (p) = 0\,,
\end{equation}
we obtain the infrared limit
\begin{equation}
\lim_{z_0 \to +\infty} W_{1/z_0} \left[ (\mu z_0)^{-\frac{\eta}{2}}
  J\right] = \frac{1}{2} \int_p \frac{J(p) 
  J(-p)}{p^2 \left(\frac{p}{\mu}\right)^{-\eta}}\,.
\end{equation}
This is obviously independent of $\ep$.  (This is in fact thanks to
the scale invariance (\ref{Wbar-scale}).)

We wish to rewrite (\ref{W-start}) in the AdS form by using the change
of variables introduced in \cite{Sathiapalan:2017frk}.  We first
rewrite
\begin{equation}
Z_{\ep, z_0} [\varphi] \equiv
\exp \left[ - \frac{1}{2} \int_p
                    \frac{\varphi (p) \varphi (-p)}{G_{z_0} (p) - G_{\ep} (p)}\right]
\end{equation}
by introducing a field $y (z,p)$ in
$D+1$ dimensions as
\begin{equation}
Z_{\ep, z_0} [\varphi] = \int \D y\Big|_{y(\ep,p) = \varphi (p)\atop y(z_0, p) =0}
\exp \left[ - \frac{1}{2} \int_\ep^{z_0} \frac{dz}{z^{D+1}} \int_p
  \frac{z \partial_z y(z,p)\cdot  z \partial_z y(z,-p)}{z^{1-D}
    \frac{\partial}{\partial z} G_z (p)}\right]\,,
\end{equation}
where the boundary values of the field $y(z,p)$ are fixed by
\begin{subequations}
\begin{align}
y (\ep, p) &= \varphi (p)\,,\\
y (z_0, p) &= 0\,.
\end{align}
\end{subequations}
Then, following 2.3 of \cite{Sathiapalan:2017frk}, we change field
variables from $y(z,p)$ to $Y(z,p)$ defined by
\begin{equation}
y (z,p) = f(z,p) Y (z,p)\,,\label{y-to-Y}
\end{equation}
where the positive function $f(z,p)$ is defined by
\begin{equation}
f(z,p)^2 \equiv z^{1-D} \partial_z G_z (p)\,.\label{f-def}
\end{equation}
(Note that the mass dimensions of $y(z,p), f(z,p), Y(z,p)$ are
$- \frac{D+2}{2}, \frac{D-2}{2}, - D$, respectively.)  Hence, we
obtain
\begin{align}
  &- \frac{1}{2} \int_\ep^{z_0} \int_p \frac{z \partial_z y(z,p) \cdot
    z \partial_z y(z,-p)}{f(z,p)^2}\nn\\
  &\quad= - \frac{1}{2} \int_\ep^{z_0} \frac{dz}{z^{D+1}} \int_p \lb
  z \partial_z Y(z,p)
  \cdot z \partial_z Y(z,-p) \right.\nn\\
&\qquad\qquad\left. + \left( \left( z \partial_z \ln f(z,p) \right)^2
    - z^{D+1} \partial_z \left( z^{1-D} \partial_z \ln f(z,p) \right)
  \right) Y(z,p) Y(z,-p)
  \rb\nn\\
  &\qquad + \frac{1}{2} \int_p \ep^{1-D} \partial_\ep \ln f(\ep, p)\,
  \cdot Y(\ep, p) Y(\ep, -p)\,,
\end{align}
where we have used $Y(z_0, p) = 0$.

We now choose $f(z,p)$ to satisfy
\begin{equation}
\left( z \partial_z \ln f(z,p) \right)^2
   - z^{D+1} \partial_z \left( z^{1-D} \partial_z \ln f(z,p) \right) =
   p^2 z^2 + \frac{m^2}{\mu^2}\,,\label{logf-equation}
\end{equation}
where $m^2$ is a squared mass parameter. This gives
\begin{align}
&- \frac{1}{2} \int_\ep^{z_0} \int_p \frac{z \partial_z y(z,p) \cdot
  z \partial_z y(z,-p)}{f(z,p)^2} \notag\\
&\quad = S_{\mathrm{AdS}; \ep, z_0} [Y] +\frac{1}{2} \int_p
\ep^{1-D} \partial_\ep \ln f(\ep, p)\, \cdot Y(\ep, p) Y(\ep, -p)\,,
\end{align}
where the action
\begin{align}
S_{\mathrm{AdS}; \ep, z_0} [Y] & 
\equiv - \frac{1}{2} \int_\ep^{z_0} \frac{dz}{z^{D+1}} \int_p \Big\lbrace
z \partial_z Y(z,p) \cdot z \partial_z Y(z,-p) \notag\\
&\qquad\qquad\qquad + \left( p^2 z^2 + \frac{m^2}{\mu^2}
            \right) Y(z,p) Y(z,-p) \Big\rbrace\label{AdSaction}
\end{align}
is defined for a massive scalar field $Y$ in the
$D+1$-dimensional AdS space with radius of curvature $\frac{1}{\mu}$.

Let us solve (\ref{logf-equation}), which amounts to
\begin{equation}
- \partial_z \left( z^{1-D} \partial_z \frac{1}{f(z,p)} \right) +
z^{1-D} \left( p^2 + \frac{m^2}{\mu^2 z^2} \right) \frac{1}{f(z,p)} = 0\,.
\end{equation}
The general solution is given by
\begin{equation}
\frac{1}{f(z,p)} = A(p) z^{\frac{D}{2}} K_\nu (p z) + B (p)
z^{\frac{D}{2}} I_\nu (p z)\,,\label{f-Bessel}
\end{equation}
where 
\begin{equation}
\nu \equiv \sqrt{\frac{m^2}{\mu^2} + \frac{D^2}{4}} > 0\,,
\end{equation}
and $I_\nu, K_\nu$ are the modified Bessel functions.  We will shortly
determine the functions of momenta $A(p), B(p)$ (both with the mass
dimension $1$) and the value of $\nu$ (equivalently $\frac{m^2}{\mu^2}$).

Now, (\ref{f-def}) and (\ref{logf-equation}) imply that $G_z
(p)/f(z,p)$ satisfies the same differential equation as $1/f(z,p)$.
Hence, we obtain
\begin{equation}
\frac{G_z (p)}{f(z,p)} = C(p) z^{\frac{D}{2}} K_\nu (pz) + D(p)
z^{\frac{D}{2}} I_\nu (pz)\,.\label{G-Bessel}
\end{equation}
(Note the mass dimension of $C, D$ is $-1$.) Moreover, (\ref{f-def}) gives
\begin{equation}
z \partial_z \frac{1}{f(z,p)} \cdot \frac{G_z (p)}{f(z,p)} -
\frac{1}{f(z,p)} z \partial_z \frac{G_z (p)}{f(z,p)} = - z^D\,.
\end{equation}
Substituting (\ref{f-Bessel}) and (\ref{G-Bessel}) into the above, and
using the Wronskian
\begin{equation}
\frac{d}{dz} I_\nu (z)\cdot K_\nu (z) - I_\nu (z) \frac{d}{dz} K_\nu
(z) = \frac{1}{z}\,,
\end{equation}
we obtain
\begin{equation}
A (p) D(p) - B (p) C (p) = 1\,.\label{ADBC}
\end{equation}

We can determine the coefficient functions $A(p)$ to $D(p)$, and the
constant $\nu$ as follows.  From (\ref{f-Bessel}) and
(\ref{G-Bessel}), we obtain
\begin{equation}
G_z (p) = \frac{C (p) K_\nu (pz) + D(p) I_\nu (pz)}{A (p) K_\nu (pz) +
  B (p) I_\nu (pz)}\,.
\end{equation}
Let us consider the limit $z \to 0+$.  We must find
\begin{equation}
\lim_{\Lambda \to +\infty} W_\Lambda \left[
  \left(\frac{\Lambda}{\mu}\right)^{\frac{\eta}{2}} J\right] = \lim_{z
  \to 0+} W_{1/z} [ (\mu z)^{-\frac{\eta}{2}} J] 
= 0\,,
\end{equation}
since this corresponds to the integration of no momentum mode.  Hence,
we obtain
\begin{equation}
\lim_{z \to +0} G_z (p) = \frac{C(p)}{A (p)} = 0\,.
\end{equation}
This gives
\begin{equation}
C (p) = 0\,.\label{C}
\end{equation}
Next, consider the limit $z \to +\infty$.  From
\begin{equation}
\lim_{z \to \infty} G_z (p) = \frac{1}{p^2
  \left(\frac{p}{\mu}\right)^{-\eta}}
\end{equation}
we obtain
\begin{equation}
\frac{D(p)}{B(p)} = \frac{1}{p^2
  \left(\frac{p}{\mu}\right)^{-\eta}}\,.\label{DB}
\end{equation}
Combining the three equations (\ref{ADBC}, \ref{C}, \ref{DB}), we
obtain
\begin{subequations}
\label{coefficients}
\begin{align}
A (p) &= \frac{1}{c}\, p \,(p/\mu)^{-\frac{\eta}{2}}\,,\\
B (p) &= c \, p\, (p/\mu)^{-\frac{\eta}{2}}\,,\\
C (p) &= 0\,,\\
D (p) &= c \,\frac{1}{p} \, (p/\mu)^{\frac{\eta}{2}}\,,
\end{align}
\end{subequations}
where we have taken $c$ to be a constant for simplicity.

To determine $c$, we must examine $f(z,p)$.  From (\ref{f-Bessel}), we obtain
\begin{equation}
\frac{1}{f(z,p)} = z^{\frac{D}{2}} p\, \left( \frac{p}{\mu} \right)^{-
  \frac{\eta}{2}} c \left( \frac{1}{c^2} K_\nu (p z) + I_\nu (p z)
\right)\,.
\end{equation}
Demanding that the change of variables from $y(z,p)$ to
$Y (z,p) = y(z,p)/f(z,p)$ be analytic at $p^2=0$, we must first choose
\begin{equation}
c^2 = \frac{\pi}{2} \frac{1}{\sin \pi \nu}
\end{equation}
so that
\begin{equation}
\frac{1}{c^2} K_\nu (p z) + I_\nu (p z) = I_{- \nu} (pz)\,.
\end{equation}
Then, since 
\begin{equation}
I_{-\nu} (z) \overset{z \to 0}{\longrightarrow} z^{-\nu} \times
\left( \frac{1}{\Gamma (1-\nu)} + \textrm{analytic in } z^2 \right)\,,
\end{equation}
we must choose
\begin{equation}
\boxed{\nu  = 1 - \frac{\eta}{2} = 1 - \gamma < 1}\label{nu-boxed}
\end{equation}
so that
\begin{equation}
  \boxed{\frac{1}{f (z,p)} = \left(\frac{\pi}{2} \frac{1}{\sin \pi
        \nu}\right)^{\frac{1}{2}} 
    z^{\frac{D}{2}} p \left(\frac{p}{\mu}\right)^{-\frac{\eta}{2}}
    I_{-\nu} (p z)} \label{f-boxed}
\end{equation}
is analytic at $p^2 = 0$.

Note that the resulting high-momentum propagator
\begin{equation}
\boxed{G_z (p) = \frac{1}{p^2 \left(\frac{p}{\mu}\right)^{-\eta}
  \frac{I_{-\nu} (pz)}{I_\nu (pz)}}}\label{Gz}
\end{equation}
is analytic at $p^2 = 0$ as long as $z$ is finite.  Only as
$z \to +\infty$, we find non-analyticity:
\begin{equation}
\lim_{z \to \infty} G_z (p) = \frac{1}{p^2
  \left(\frac{p}{\mu}\right)^{-\eta}}\,.
\end{equation}

To summarize so far, for the choice of the high-momentum propagator
(\ref{Gz}), we can rewrite (\ref{W-start}) as
\begin{align}
  e^{W_{1/z_0} [J]} &= \int [d \varphi] \exp \left[ \frac{1}{2} \int_p J(p)
    J(-p) G_\ep (p) \right]\nn\\
  &\quad \times \exp \left( \int_p J(p) \varphi (-p) + \frac{1}{2} \ep^{-D}
    \int_p \ep \partial_\ep \ln f(\ep,p) \cdot \frac{1}{f(\ep,p)^2}
    \, \varphi (p) \varphi (-p)
  \right)\nn\\
  &\quad \times \int \D Y\Big|_{Y(\ep, p) = \varphi (p)/f(\ep, p)\atop
    Y(z_0,p) = 0} \,e^{S_{\mathrm{AdS}; \ep, z_0}
    [Y]}\quad, \label{AdS-zzero}
\end{align}
where the AdS action is given by (\ref{AdSaction}).  By construction,
(\ref{AdS-zzero}) is independent of $\ep$, but it depends on $z_0$.

In the limit $z_0 \to +\infty$, we obtain the generating functional:
\begin{align}
e^{\W [J]} &= \int [d \varphi] \exp \left[ \frac{1}{2} \int_p J(p) J(-p) G_\ep (p)
                    \right]\nn\\
&\quad \times \exp \left( \int_p J(p) \varphi (-p)
                    + \frac{1}{2} \ep^{-D} \int_p
                    \ep \partial_\ep \ln f(\ep,p) \cdot
  \frac{1}{f(\ep,p)^2} \,\varphi (p) \varphi (-p) \right)\nn\\
&\quad \times \int \D Y\Big|_{Y(\ep, p) = \varphi (p)/f(\ep, p)\atop
Y(+\infty,p) = 0} \,e^{S_{\mathrm{AdS}; \ep, \infty} [Y]}\label{AdS-form}
\end{align}
Since this is independent of $\ep$, we can also take the limit $\ep
\to 0+$.   Using
\begin{align}
\frac{1}{f(\ep, p)} &\longrightarrow \left(\frac{\pi}{2} \frac{1}{\sin \pi
                      \nu}\right)^{\frac{1}{2}} \ep^{\frac{D}{2}-1}  (\mu
                      \ep)^\gamma \frac{1}{\Gamma (1-\nu)} \propto
                      \ep^{\frac{D}{2}-1} (\ep \mu)^\gamma \,,\\
G_\ep (p) &\longrightarrow \ep^2 \left(\mu \ep\right)^{-\eta}\frac{\Gamma
            (1-\nu)}{\Gamma (1+\nu)} \propto \ep^2 (\ep \mu)^{-\eta}\,,
\end{align}
we obtain, for $\ep \to 0+$,
\begin{align}
  e^{\W [J]} &= \int [d \varphi] \exp \left( \int_p J(p) \varphi (-p) -
    \frac{\Delta_-}{2} \ep^{-2} (\ep \mu)^\eta \int_p
    \alpha^2 \varphi (p) \varphi (-p) \right)\nn\\
  &\quad \times \int \D Y\Big|_{Y(\ep, p) = \alpha \mu^\gamma
    \ep^{\Delta_-} \varphi (p)\atop Y(z_0,p) = 0} \,e^{S_{\mathrm{AdS};
      \ep, \infty} [Y]}\,,\label{AdS-form}
\end{align}
where
\begin{align}
\Delta_- &\equiv \frac{D}{2} - \nu = \frac{D-2}{2} + \gamma\,,\\
\alpha &\equiv \frac{1}{\sqrt{2}} \left(\frac{\Gamma (\nu)}{\Gamma
         (1-\nu)}\right)^{\frac{1}{2}}\,.
\end{align}

In the literature it may be more common to write $\alpha \varphi (p)$ as $\varphi
(p)$ and $\frac{1}{\alpha} J(p)$ as $J(p)$ so that
\begin{align}
e^{\W [J]} &= \int [d \varphi]  \exp \left( \int_p J(p) \varphi (-p)
                    - \frac{\Delta_-}{2} \ep^{-2} (\ep \mu)^\eta \int_p
             \varphi (p) \varphi (-p) \right)\nn\\
&\quad \times \int \D Y\Big|_{Y(\ep, p) = \mu^\gamma \ep^{\Delta_-} \varphi (p)\atop
Y(z_0,p) = 0} \,e^{S_{\mathrm{AdS}; \ep, \infty} [Y]}\,.\label{AdS-form}
\end{align}
We then obtain
\begin{equation}
\W [J] = \frac{1}{2} \int_p \frac{\alpha^2}{p^2
  \left(\frac{p}{\mu}\right)^{-\eta}}\,J(p) J(-p)\,.
\end{equation}
This reproduces the prescription of the alternative quantization of
the AdS/CFT correspondence \cite{Klebanov:1999tb} (reviewed nicely in
Appendix of \cite{Faulkner:2010jy}) for computing the two-point
function.

In \cite{Sathiapalan:2017frk} it was pointed out that when the ERG
equation is mapped to AdS space there remains a boundary term
depending on the function $f(p)$.  We see this in \eqref{AdS-form}
also. But note that it is analytic in $p$ and therefore does not
affect the all important non-analytic piece.

\section{Discussion}\label{section:discussion}

We have managed to derive the AdS/CFT correspondence from ERG, but our
derivation is not without faults.  We discuss three issues here.

\subsection{Non-analyticity of the Wilson action at zero momentum}

As we have already pointed out in Sec. \ref{section:background}, our
Wilson action (\ref{quad-action}) is not analytic at $p^2=0$ (hence
non-local) due to the anomalous dimension $\eta$.  This is partially
because we are treating a composite field as an elementary field
$\phi$.  Even in the free massless theory in $D$ dimensions, the
composite field $\phi^2$ has scale dimension $D-2$ so that its
anomalous dimension is $(D-2)/2$ compared with the canonical scale
dimension $(D-2)/2$ of $\phi$.

Another reason for the non-analyticity is that we are not taking
interactions into account.  Consider a Wilson action whose quadratic
part is given by
\begin{equation}
  \SL = - \frac{1}{2} \int_p \frac{p^2}{K(p/\Lambda)}
  \frac{1}{1 + K(p/\Lambda) \left(
      \left(\frac{p^2+m_\Lambda^2}{\mu^2}\right)^{\frac{\eta}{2}} - 1
    \right)} \phi (p) \phi (-p) \,,
\label{action-with-mass}
\end{equation}
where the cutoff dependence of the squared mass $m_\Lambda^2$ is
determined by four-point interactions.  (The cutoff dependence of the
four-point interactions is determined by the sixth point interactions,
and so on.)  The squared mass restores the analyticity of the action
at $p^2=0$.  The corresponding generating functional is
\begin{equation}
W_\Lambda [J] = \frac{1}{2} \int_p \frac{J (p) J(-p)}{p^2
  \left(\frac{p^2 + m_\Lambda^2}{\mu^2}\right)^{\frac{\eta}{2}}+
  R_\Lambda (p)} +
\textrm{quartic and higher}\,,
\end{equation}
and it reproduces the same two-point function in the limit $\Lambda
\to 0+$, if we assume
\begin{equation}
\lim_{\Lambda \to 0+} m_\Lambda^2 = 0\,.
\end{equation}

\subsection{Non-analyticity of the cutoff function $K$}

In the ERG formulation, the choice of a cutoff function $K(p/\Lambda)$
is totally arbitrary as long as it satisfies $K(0)=1$, and it
decreases rapidly for $p > \Lambda$.  But we usually assume
$K(p/\Lambda)$ to be analytic at $p^2 = 0$:
\begin{equation}
K(p/\Lambda) = 1 + \textrm{integral powers of $p^2$}\,.
\end{equation}

Now, in Sec. \ref{section:derivation} we have chosen a particular cutoff
function so that
\begin{equation}
G_{1/\Lambda} (p) \equiv \frac{1}{p^2
  \left(\frac{p}{\mu}\right)^{-\eta} +
  \left(\frac{\Lambda}{\mu}\right)^{-\eta} R_\Lambda (p)} = \frac{1}{p^2
  \left(\frac{p}{\mu}\right)^{-\eta} \frac{I_{-\nu} (p/\Lambda)}{I_\nu
    (p/\Lambda)}}\,.
\end{equation}
This is obtained from (\ref{Gz}).  This implies
\begin{equation}
R_\Lambda (p) = p^2 \left(\frac{p}{\Lambda}\right)^{-\eta} \left( \frac{I_{-\nu}
      (p/\Lambda)}{I_{\nu} (p/\Lambda)} - 1 \right)\,.
\end{equation}
(Note that this has the form (\ref{RLambda-dep}).)  We then obtain,
from (\ref{RLambda}),
\begin{equation}
K(p/\Lambda) = \frac{1}{1 + \frac{\left(\frac{p}{\Lambda}\right)^\eta
    \frac{I_\nu (p/\Lambda)}{I_{-\nu} (p/\Lambda)}}{1 -  \frac{I_\nu
      (p/\Lambda)}{I_{-\nu} (p/\Lambda)}}}\,,
\end{equation}
where 
\begin{equation}
  \left(\frac{p}{\Lambda}\right)^\eta \frac{I_\nu
    (p/\Lambda)}{I_{-\nu} (p/\Lambda)} = \mathrm{O} (p^2/\Lambda^2)
\end{equation}
is analytic at $p^2=0$ since $2 \nu = 2 - \eta$.  Therefore, we find
\begin{equation}
K(p/\Lambda) = 1 + \mathrm{const} \, (p/\Lambda)^2 + \mathrm{const}\,
(p/\Lambda)^{4 -\eta} + \cdots\,.
\end{equation}
This is not analytic.  Since no physics depends on the choice of
$K(p/\Lambda)$, one could argue, perhaps, that using a cutoff function
non-analytic at $p^2=0$ is acceptable.

\subsection{Absence of interactions}

We have already discussed the importance of introducing interactions
to restore the analyticity or locality of the Wilson action.  From the
ERG perspectives, it is natural to introduce interactions only at the
boundary $z = \ep$ of the AdS space.  Whether or not interactions are
induced in the bulk of the AdS space is left for a future study. We
make some preliminary remarks on this in Section VI.

\section{Identifying the Anomalous Dimension from the ERG equation}\label{section:identification}
\label{elucidation}

As mentioned in Section II C, in \cite{Sathiapalan:2017frk} the
starting point of the discussion was Polchinski's ERG without any
parameter for anomalous dimension. In the AdS version of the evolution
operator, however, a new parameter, absent in the original Polchinski
evolution operator, made its appearance.  This parameter is contained
in the specification of the function $f(z,p)$ used in the change of
variables. This is the parameter $\nu$ and becomes the (anomalous)
dimension of the boundary operator as seen from the two point function
calculation.  In the AdS scalar field equation it shows up as the mass
of the scalar field.  In this paper we have started with a modified
Polchinski ERG {\em with} a parameter $\nu$, for anomalous dimension.
$\nu$ was chosen so that the correct high energy propagator $G_z$ of
the fixed point theory with anomalous dimension is reproduced.  In
this section we would like to elucidate the role of this parameter
from the point of view of the ERG equation and explain how a choice of
variables dictates the equation.

As pointed out originally by Wilson and others, (see for instance,
\cite{Wilson:1973jj} or Bell and Wilson,
\cite{Bell:1975wtp,Bell:1974vv}), there are two important aspects in
an RG --- one is the coarse graining and the other is a rescaling of
the field variables. In \cite{Bell:1975wtp,Bell:1974vv}, they
exemplify this with the following simple transformation, $T$ (in their
notation): 
\be \label{WB} T e^{-H[S]}=\int _\sigma e^{-\hf a \int _q
  (S_q-b\sigma_{q/2})^2}e^{-H[\sigma]} .
\ee
The initial field variable is $\sigma$ and the final field variable if
$S$. The subscript on the field variable has changed from $q/2$ to
$q$. $q$ is dimensionless and this corresponds to a change of the
cutoff from $\lm\to \lm/2$. This is the coarse graining. The factor
$b$ is the rescaling. After $n$ steps the rescaling becomes $b^n$. The
factor $a$ can be changed by a scaling of {\em both} $\sigma$ and $S$,
and has no effect on the physics.

In a field theory the field variables are integrated over. Thus a
constant rescaling does not change the physics. Thus in the above
example, $H[S]$ and $H[bS]$, even though they have different
mathematical forms, describe the same physics in the sense that one
can map correlations calculated with one variable to correlations
calculated with the other variable by a change of normalization. The
S-matrix is in fact invariant. Thus for instance if $H[S]$ describes a
scale invariant theory (i.e. a fixed point) then so does $H[bS]$. Thus
if an {\em exact} RG transformation \footnote{If the RG is not exact,
  the choice of variable becomes important --- because a bad choice of
  variable throws out relevant information. But in an exact RG no
  information is lost.} is done on a critical physical system in the
basin of attraction of a fixed point, it will move towards the fixed
point regardless of the choice of variables --- this is a generic
property of all physical systems as we flow to the IR. This does not
mean that the Hamiltonian is mathematically form invariant at the
fixed point. {\em A fixed point Hamiltonian is an equivalence class of
  mathematical expressions for the Hamiltonian related by rescalings
  of the field variable.} The physical property that defines a fixed
point is that physical quantities such as the S-matrix will have a
scale invariance --- absence of any characteristic scale. All this
goes to show that the choice of $b$ does not change the physics.

However, if one wants to make the Hamiltonian {\em mathematically form
  invariant} under the ERG, then a particular rescaling parameter ($b$
in the above example) has to be chosen --- the precise value depends on
the details of the interacting theory. This is a choice of field
normalization. A natural choice is to make sure that the kinetic term
has a fixed normalization. If this normalization is implemented then
the fixed point Hamiltonian is mathematically identical after the
transformation. This is convenient in an actual calculation because
requiring that the Hamiltonian be mathematically identical after an RG
transformation leads to a well defined mathematical ``fixed point
equation''. \footnote{If the correct $b$ is not chosen then when
  approaching a fixed point, after each RG iteration, the
  normalization of the kinetic term will get multiplied by a constant
  factor (not equal to 1) and this can lead to complications.}

Let us now go to the connection between $b$ and the anomalous
dimension. In the fixed point equation one works with dimensionless
variables. Thus the variable $\sigma$ has to be scaled by a factor to
undo the change of scale from $\lm\to \lm/2$ and keep the
normalization of the kinetic term fixed. In a free theory this is
given by a factor $2^{-\frac{(D+2)}{2}}$ which reflects the
engineering dimension of the field.  So we get
\[
b=  2^{-\frac{(D+2)}{2}}.
\]  
In an interacting theory there are further contributions at each
iteration and this modifies the scaling dimension by adding an
anomalous dimension to the engineering dimension and gives
\[
b=   2^{-\frac{(D+2+\eta)}{2}}.
\]  

Let us transcribe this to continuous time ERG where $\lm (t) = \lm(0)
e^{-t}$. A factor $b$ becomes $e^{-\frac{(D+2+\eta)}{2}t}$. If we
denote by $x_i$ the initial variable ($\sigma$ in the above example)
and $x_f$ the final variable ($S$ in the above example) then the
transformation \eqref{WB} can be written in the form of a time
dependent rescaling as (the notation used in
\cite{Sathiapalan:2017frk} where momentum labels are suppressed)
\be \label{G3} \psi(x_f,t_f)=\int _{x_i} e^{-\hf A(e^{\alpha t_f}x_f
  -e^{\alpha t_i} x_i)^2}\psi(x_i,t_i) .
\ee
The conclusion we reach is that to determine the anomalous dimension
of the evolution equation we can write the integrating kernel in the
form \eqref{G3} and read off the anomalous dimension $\eta$ from
$e^{\alpha t}$.\footnote{As mentioned above the value of $\eta$ is a
  property of the action and should be chosen depending on the action
  one starts with.}  In Appendix \ref{Appendix A} we illustrate this
point with two examples of standard ERG equations.

This also suggests that if we do a field redefinition involving a time
dependent rescaling, of the form $x(t) = y(t)e^{\mu t}$, the scaling
dimension is changed by an amount $\mu$. Thus an ERG with anomalous
dimension can be related to one without anomalous dimension, by such a
field redefinition. If one wants a mathematical fixed point the
anomalous dimension has to be chosen correctly. The precise value will
depend on the interactions.

This then answers the question raised at the beginning of this
section: In \cite{Sathiapalan:2017frk} the starting point was
Polchinski's ERG without anomalous dimension. The integrating kernel
is of the form 
\be 
\label{G} \psi(x_f,t_f)=\int _{x_i} e^{-\hf
  \frac{(x_f - x_i)^2}{G_i-G_f}}\psi(x_i,t_i) .
\ee
With the change of
variables $x=fy$ we obtain 
\be 
\label{G1} \psi(y_f,t_f)=\int _{y_i}
e^{-\hf \frac{(y_f f(t_f)- y_if(t_i))^2}{G_i-G_f}}\psi(y_i,t_i) .
\ee

We have seen that $f\approx e^{-(\frac{D-2+\eta)}{2}t}$. Clearly the
ERG equation obeyed in the new variables will have an anomalous
dimension parameter.  This explains the appearance of this parameter
in the AdS equation in \cite{Sathiapalan:2017frk}.  In
Sec. \ref{section:background} C we have already seen how
(\ref{ERGdiffeqWtilde}) is related to (\ref{ERGdiffeqW}) by change of
variables.

In this section we have shown the role of field redefinitions (or
``wave function renormalization'' in perturbative calculations) in
introducing anomalous dimension in an ERG equation. This is important
for locating the mathematical fixed point of the equation. We have
also seen that the dimension can be read off from the integral
formula. Some other examples of these are given in Appendix A.

\section{Nontrivial Fixed Point Action}
\label{nonlinear}

In this section we consider a nontrivial fixed point action.  To
begin with we use the usual Polchinski ERG formalism. The kinetic term
is $\hf x^2 G^{-1}$ and the interacting part is $S_0$.

\[
S_{FP}= \hf x^2 G^{-1} + S_0(x)\,.
\]
Let the perturbation be $S_1$ so that the full action is
\[
S= \hf x^2 G^{-1} + S_0(x) + S_1 (x)\,.
\]


Then in our earlier notation, the ``wave functions'' are given by
\[
\psi = e^{-S}=e^{-[\hf x^2 G^{-1} + S_0(x,t) + S_1 (x,t)]}\,,\qquad
\psi'=e^{-[ S_0(x,t) + S_1 (x,t)]}\,. 
\]
Polchinski's equation is
\be	\label{101}
 {\p \psi'\over \p t} = -\hf \dot G {\p^2 \psi'\over \p x^2}\,.
\ee
What is special is that $S_0$ by itself satisfies Polchinski's
equation --- eventually it will be taken to be a fixed point solution.
Thus we have the following two equations: 
\be \label{s0} 
{\p S_0\over
  \p t} =\hf \dot G [-{\p^2S_0\over \p x^2} + ({\p S_0\over \p x})^2 ]
\ee
and 
\be \label{s1} 
{\p S_0\over \p t} +{\p S_1\over \p t}=\hf \dot
G [-{\p^2S_0\over \p x^2} + ({\p S_0\over \p x})^2 - {\p^2S_1\over \p
  x^2} + ({\p S_1\over \p x})^2 + 2({\p S_0\over \p x}) ({\p S_1\over
  \p x})] \,.
\ee
Subtracting \eqref{s0} from \eqref{s1} we get
\be \label{s2}
{\p S_1\over \p t}=\hf \dot G [- {\p^2S_1\over \p x^2} + ({\p S_1\over
  \p x})^2 + 2({\p S_0\over \p x}) ({\p S_1\over \p x})]\,. 
\ee
Since $S_0$ is a solution of \eqref{s0}, its form is ({\em in
  principle}) known as a function of time. In the case that $S_0$ is
chosen to be a fixed point solution, its time dependence can be
specified very easily: expressed in terms of {\em rescaled and
  dimensionless} variables it has no time dependence. This is
equivalent to saying that the dimensionless couplings are constant in
RG-time, $t$, i.e., they have vanishing beta functions. One can work
backwards and determine the exact $t$-dependence in terms of the
original variables. 
\eqref{s2} can be used to define a modified Hamiltonian evolution
equation for the wave function $\psi''=e^{-S_1(x,t)}$:
\be \label{s3}
{\p \over \p t}\psi '' = -\hf \dot G [{\p ^2\over \p x^2} - 2 ({\p S_0\over \p x}){\p \over \p x}]\psi ''\,.
\ee
Note that the term involving $S_0$ is like a gauge field coupling - in
fact it is ``pure gauge''.  The Action functional corresponding to
this Hamiltonian is derived in Appendix B using canonical
methods. While it involves more algebra, it can be applied even when
$S_0$ is not a solution of the ERG equation.
 
In the end the result can be summarized very simply. Start with the
usual RG evolution of $\psi'$:
\be
e^{-S_0\left(x(t_f),t_f\right)-S_1\left(x(t_f),t_f\right)}=\int {\cal D}xe^{\hf
  \int_{t_i}^{t_f} dt~\hf \frac{1}{\dot G}(\frac
  {dx}{dt})^2}e^{-S_0\left(x(t_i),t_i\right)-S_1\left(x(t_i),t_i\right)} \,.
\ee
Take
$S_0(x(t_f)),t_f)$ into the RHS to get 
\be \label{eo} e^{-S_1\left(x(t_f),t_f\right)}=\int {\cal D}x\, e^{\hf
  \int_{t_i}^{t_f} dt~\hf \frac{1}{\dot G}(\frac
  {dx}{dt})^2}
e^{S_0\left(x(t_f),t_f\right)-S_0 \left(x(t_i),t_i\right)-S_1
  \left(x(t_i),t_i\right)} 
\,. 
\ee
Introduce the evolution of $S_0$ into the functional integral by
writing it as an integral of a total derivative:
\[
\Delta S_0= S_0(x(t_f),t_f)-S_0(x(t_i),t_i) = \int_{t_i}^{t_f}
dt~\frac{dS_0(x(t),t)}{dt}
\]
\be 
=\int _{t_i}^{t_f} dt~[\frac{\p S_0(x(t),t)}{\p t}+\frac{
  dx(t)}{dt} {\p S_{0}(x(t))\over \p x(t)}] \ee
to get
\be  \label{eo1}
e^{-S_1 \left(x(t_f),t_f \right)}=\int {\cal D}x\,e^{\hf
  \int_{t_i}^{t_f} dt~\left[ \hf
  \frac{1}{\dot G}(\frac {dx}{dt})^2+ \frac{\p S_0\left(x(t),t\right)}{\p
    t}+\frac{ dx(t)}{dt} {\p S_{0}\left(x(t), t\right)\over \p
    x(t)}\right]}e^{-S_1\left(x(t_i),t_i\right)} \,.
\ee 
Thus the evolution of the perturbation involves an evolution operator
that has a non linear term, but is in fact a {\em total derivative}.

The total derivative can also be rewritten in other forms using the
ERG for $S_0$. Thus for instance (using field theory notation) it can
be rewritten as (see \eqref{Appendix B}):
\[
  S_{bulk}[\phi(p,t)]=\hf\int dt ~\int {d^Dp\over (2\pi)^D}~
  \left[-{1\over \dot G(p)} {\partial \phi(p,t)\over \partial
        t}{\partial \phi(-p,t)\over \partial t} \right.
\]
\be	\label{action}
\left.- \dot G(p) \left\lbrace{\delta
  S_0[\phi(p,t),t]\over \delta \phi(p,t)}{\delta S_0[\phi(p,t),t]\over \delta
  \phi(-p,t)} -{\delta^2 S_0[\phi(p,t),t]\over \delta
  \phi(-p,t)\delta
  \phi(p,t)}\right\rbrace - \frac{\delta S_0[\phi(p,t),t]}{\delta
  \phi(p,t)}\frac{\p \phi(p,t)}{\p t}\right]\,.  
\ee
%

%
%

What we have here is a nonlinear action describing ERG evolution ---
because of the presence of $S_0[\phi]$ in the action, it is no
longer quadratic in $\phi$. This is unusual in ERG literature:
Acting on $\psi=e^{-S_B}$, the evolution given by
  Polchinski ERG equation \eqref{101}, is linear.  (The equation is of
  course nonlinear when written in terms of $S_B$.)

On the other hand in discussions of holographic RG (see for instance
\cite{Heemskerk:2010hk,Faulkner:2010jy}) the Hamiltonian radial
evolution of the boundary action, $S_B[\phi]$, is given by
\be \label{nl}
\p_\eps S_B =\int_{z=\eps}d^dx~\{(\frac{\dd S_B}{\dd \phi})^2 +\hf
\p_\mu\phi\p^\mu\phi+V[\phi]\} \,.
\ee 
Here $V[\phi]$ is the potential for
the scalar field in the bulk action. Thus the holographic RG evolution
in \eqref{nl} is nonlinear in $\phi$ even when acting on
$\psi=e^{-S_B[\phi]}$ due to the presence of $V[\phi]$ .  Indeed in
AdS/CFT correspondence, $V[\phi]$ plays an important role in
determining higher point correlation functions and also beta functions
(see for example \cite{Witten:1998qj, Bzowski:2015pba}) of the boundary
theory through ``Witten diagrams".  It is after all in this sense that
the bulk action has information about the boundary theory.

In our discussion in this section, nonlinear terms have  appeared,  but
 whether the nonlinearity introduced here is of the same
ilk as in \eqref{nl} is not clear to us. In particular it holds here that if $S_0$
satisfies the RG equation by itself, then this nonlinear piece is a
total derivative. The origin of nonlinear terms in holographic RG
from the viewpoint of ERG is thus an open question.

\section{Conclusion}\label{section:conclusion}

Some aspects of the mapping of ERG equation to a holographic AdS form
that was introduced in \cite{Sathiapalan:2017frk} were clarified in
this paper with a view to extending the results to nontrivial fixed
point theories. The ERG written there was for the Wilson action. The
bulk AdS field corresponds to a field in the boundary theory rather
than a source. This then corresponds to what is called ``alternative
quantization''. The main point of this paper is to clarify how the map
from ERG to AdS can be done in the presence of an anomalous dimension
parameter. This was done in a concrete example where the generating
functional is quadratic but has an anomalous dimension parameter.  In
general one expects anomalous dimension when there are
interactions. Here the focus is on the mapping to AdS, so we simplify
calculations by introducing the anomalous dimension into the two point
function by hand. The same techniques should work in more realistic
theories also. The results reproduce those obtained in the literature
using the AdS/CFT conjecture.

%

Thus many of the results of AdS/CFT calculations for two point
functions are obtained directly starting from the ERG of the boundary
CFT without invoking any conjecture. Also string theory itself plays
no direct role at this level. As mentioned in
\cite{Sathiapalan:2017frk} string theory possibly plays a role in
making the bulk theory well defined as a quantum theory.

There are several other open questions. We have studied only the
propagator of this theory. Higher order correlators are hard to study
because we are dealing with a QFT. In order to obtain concrete results 
an expansion parameter such as 1/N is needed.

This paper discusses only perturbations involving an elementary scalar
field.  Equally interesting are questions involving composite
fields. This was also considered in a general way in
\cite{Sathiapalan:2017frk}.  Once again a large N expansion is
required to do these computations.

The role of dynamical gravity has not been discussed thus far. This
needs to be addressed.

\appendix

\section{Other ERG equations}
\label{Appendix A}

Let us study two of the standard ERG equations: 
We let our Wilson action be (using the simplified notation suppressing momentum labels)
\[
S= \hf x G^{-1}x + S^I_\lm
\]
with $G$ the  propagator, $G= \frac{K(\lm)}{p^2}$. We take $G_0 = \frac{K(\lo)}{p^2}=\frac{K_0}{p^2}$. (We take $K_0=1$ for convenience.)

Polchinski's equation with anomalous dimension $\etat$ in simplified
notation, with momentum dependence suppressed is:
\be \label{H}
\frac{\p \psi }{\p t}=[-\hf( \dot G+\eta \frac{G(G_0-G)}{G_0})
\frac{\p^2}{\p x^2} - (\frac{\dot G}{G}+\etat) x \frac{\p}{\p
  x}]\psi\,.
\ee

Another ERG equation with anomalous dimension was written down in 
 \cite{Ball:1994ji},
 \cite{Osborn:2011kw} 
 which, in simplified notation, is:  
\be   \label{H1}
 \frac{\p \psi}{\p t} =
[- \hf \dot G  \frac{\p^2 }{\p x^2}  -(\frac{\dot G}{G}+\etat) x
\frac{\p }{\p x}]\psi \,.
\ee

The integral evolution operator for both these have the same form:
\be   \label{wa5}
  \psi(x_f,t_f)=
  \int dx_ie^{-\hf \frac{(\frac{x_fe^{-\etat
          t_f}}{G_f}-\frac{x_ie^{-\etat t_i}}{ G_i})^2}{H_f^{-1}-
      H_i^{-1}}}\psi(x_i,t_i) \,.
\ee

Comparing with the form \eqref{G} we see that $e^{\etat(t_f-t_i)}$
gives the time dependent relative scaling between $x_f$ and $x_i$
corresponding to anomalous dimension $\etat$.

The equations  differ in the form of $H$. Polchinski's equation gives
 \be	\label{h}
 H^{-1}=(\frac{1}{G}-\frac{1}{G_0})e^{-\eta t}
 \ee
 and \eqref{H1}:
\be	\label{h1}
\frac{d H^{-1}}{dt}=\frac{d (G^{-1})}{dt}e^{-\eta t}\,.
\ee
     
We now put back the momentum dependence and go back to more standard notation:
\be \label{green}
G= \frac{K}{p^2},~~~G_0=\frac{1}{p^2},~~~~~~~x_i=\phi_i(p),
~~~x_f=\phi_f(p)\,. 
\ee
In the Polchinski case  from \eqref{h}
\[
 H=\frac{Re^{\eta t}}{p^4},  ~~~~~~~~       R =
 (\frac{p^2K}{1-K}),~~~e^{\eta t}=(\frac{\Lambda}{\mu})^{-\eta} 
\]
we get Polchinski's equation with anomalous dimension by substituting
these in \eqref{H}: 
\begin{align}
- \Lambda \frac{\partial}{\partial \Lambda} e^{\SL [\phi]} 
&= \int_p \left[  \left( \frac{\Delta (p/\Lambda)}{K(p/\Lambda)} -
  \etat \right) \phi (p) \frac{\delta}{\delta \phi
  (p)}\right.\notag\\
&\quad\left. + \frac{1}{p^2} \lb \Delta (p/\Lambda) - \eta
  K(p/\Lambda) \left(1 - K(p/\Lambda)\right)\rb \frac{1}{2}
  \frac{\delta^2}{\delta \phi (p) \delta \phi (-p)} \right] \, e^{\SL
  [\phi]}\,.
\end{align}
For $\Lambda_2 < \Lambda_1$, this is solved by
the kernel substituting in \eqref{wa5}:
\begin{equation}
e^{S_{\Lambda_2} [\phi]} 
= \int [d\phi'] \, \mathcal{P}_{\Lambda_2, \Lambda_1} [\phi, \phi'] \,
e^{S_{\Lambda_1} [\phi']}\,,
\end{equation}
where the evolution operator is given by
\begin{align} 
&\mathcal{P}_{\Lambda_2, \Lambda_1} [\phi, \phi']
\equiv \exp \left[ - \frac{1}{2} \int_p
  \frac{1}{\left(\frac{\Lambda_2}{\mu}\right)^\eta
    \frac{1}{R_{\Lambda_2} (p)} - \left(\frac{\Lambda_1}{\mu}\right)^\eta
    \frac{1}{R_{\Lambda_1} (p)} }\right.\notag\\
&\left.\quad \times \lb
\left(\frac{\Lambda_1}{\mu}\right)^{\frac{\eta}{2}} \frac{\phi'
  (p)}{K(p/\Lambda_1)} -
  \left(\frac{\Lambda_2}{\mu}\right)^{\frac{\eta}{2}} \frac{\phi 
  (p)}{K(p/\Lambda_2)} \rb
 \lb
\left(\frac{\Lambda_1}{\mu}\right)^{\frac{\eta}{2}} \frac{\phi'
  (-p)}{K(p/\Lambda_1)} -
  \left(\frac{\Lambda_2}{\mu}\right)^{\frac{\eta}{2}} \frac{\phi 
  (-p)}{K(p/\Lambda_2)} \rb \right]\,.\label{evolution-eta}
\end{align}

Similarly \eqref{H1} \cite{Ball:1994ji}, \cite{Osborn:2011kw}, can be
obtained by the same substitution.
\begin{align}
- \Lambda \frac{\partial}{\partial \Lambda} e^{\SL [\phi]} 
&= \int_p \left[  \left( \frac{\Delta (p/\Lambda)}{K(p/\Lambda)} -
  \gamma \right) \phi (p) \frac{\delta}{\delta \phi
  (p)}\right.\notag\\
&\quad\left. + \frac{1}{p^2} \lb \Delta (p/\Lambda) 
   \rb \frac{1}{2}
  \frac{\delta^2}{\delta \phi (p) \delta \phi (-p)} \right] \, e^{\SL
  [\phi]}\,.
\end{align}
The integrating kernel remains the same as in \eqref{evolution-eta}, 
but $R$ is defined by the  more complicated relation
\[
e^{-\eta t}\frac{d}{dt}(\frac{p^4}{R}) = \frac{d}{dt}
(\frac{1-K}{p^2K}e^{-\eta t})=\frac{d}{dt}(H^{-1}) \,.
\]

In each case we see that the anomalous dimension can be read off from
the powers of $\lm$ multiplying the fields in the exponent in
\eqref{evolution-eta}.

\section{Action for the Nontrivial Fixed Point Hamiltonian}
\label{Appendix B}

In this appendix we derive \eqref{action} using Hamiltonian methods.

Our starting point is 
\be 
{\p \over \p t}\psi '' = -\hf \dot G \left[{\p ^2\over \p x^2} - 2
  ({\p S_0\over \p x}){\p \over \p x}\right]\psi ''\,. 
\ee
Note that in principle one can postulate this as an ERG equation even if $S_0$ is not a solution of Polchinski's ERG equation as assumed in Section VI. To that extant the derivation in this Appendix is more general.

We rotate to Minkowski space: $it_M=t_E$.
Hence,
\[
e^{iS_M}=e^{i\int dt_M (-V)} =e^{-\int dt_E V}=e^{-S_E}\,.
\]
Thus, $S_E=-iS_M$.  Thus in the above case  $S_{0}=-iS_{0M}$.
Let $-\tau_E=G$ so that
\[
\frac{1}{\dot G}\frac{\p}{\p t_E} = \frac{\p}{\p G}=-\frac{\p}{\p \tau_E}\,.
\]

Write \eqref{s3} as
\[
{\p\over \p t_E} =-\hf \dot G \left[{\p ^2\over \p x^2} - 2 ({\p S_0\over
    \p x}){\p \over \p x}\right]\,.
\]
This implies
\begin{align*}
{\p\over \p G}&=-{\p\over \p \tau_E}=i{\p\over \p \tau_M}\\
              &=-\hf \left[{\p^2\over \p x^2} - 2 ({\p S_0\over \p x}){\p
                \over \p x}\right] =-\hf \left[{\p^2\over \p x^2} - 2
                ({-i\p S_M\over \p x}){\p \over \p x}\right]\\ 
& = \hf \left[-{\p^2\over \p x^2} +2 ({\p S_M\over \p x})(-i{\p \over
  \p x})\right] 
=\hf \Big[P^2 + 2 \underbrace{({\p S_M\over \p x})}_{A_x}P\Big]\,.
\end{align*}

Writing in terms of $P= -i{\p\over \p x}$, we get finally for the
Hamiltonian in Minkowski spacetime
\begin{align}
H &= \hf[P^2 + 2 {\p S_{M}\over \p x}P]=\hf[P^2 +PA_x+A_x P + i
    \frac{\p A_x}{\p x}] \notag\\
&=\hf [P^2+PA_x+A_xP +A_x^2] + \hf i \frac{\p A_x}{\p x}-\hf A_x^2\notag\\
&=\hf (P+A_x)^2+\hf i \frac{\p A_x}{\p x}-\hf A_x^2\,.
\end{align}
This gives 
\[
\dot x  =P+A_x=P + {\p S_{M}\over \p x}\,,
\]
and one can obtain using $L=x\dot P -H$ an action
\[
L=\hf \dot x^2 - \dot x A_x + \hf A_x^2 - \hf i \frac{\p A_x}{\p x}\,.
\]
Here $A_x(x)$ is to be understood as $A_x(x(t))$, and
$\frac{\p A_x}{\p x}=\frac{\p A_x(x(t))}{\p x(t)}$.  In Minkowski
spacetime we obtain
\begin{align*}
iA_x &= i {\p S_{M}\over \p x} \implies i \frac{\p A_x}{\p x}= i
       \frac{\p^2S_M}{\p x^2}\,, \\ 
L &=\hf \dot x^2 - \dot x {\p S_{M}\over \p x}+ \hf ({\p S_{M}\over \p
    x})^2 - \hf i \frac{\p^2S_M}{\p x^2}\,,\\ 
iS_M &= i\int d\tau_M~[\hf \dot x^2 - \dot x {\p S_{M}\over \p x}+ \hf
       ({\p S_{M}\over \p x})^2 - \hf i \frac{\p^2S_M}{\p x^2}]\,. 
\end{align*}
In Euclidean space this becomes
\[
-S_E=\int d\tau_E~[-\hf \dot x^2 - \frac{ dx}{d\tau_E} {i\p S_{M}\over
  \p x}+ \hf ({\p S_{M}\over \p x})^2 - \hf i \frac{\p^2S_M}{\p
  x^2}]\,. 
\]
Hence, 
\[
S_E=\int d\tau_E~[\hf \dot x^2 -\frac{ dx}{d\tau_E} {\p S_{0}\over \p
  x}+\hf ({\p S_{0}\over \p x})^2 - \hf  \frac{\p^2S_0}{\p x^2}]\,. 
\]

Now reintroduce $G$ by $dG= dt_E \frac{dG}{dt_E}=dt_E \dot G$:

{\bf i)}
\[
\hf \int d\tau_E (\frac {dx}{d\tau_E})^2= -\hf \int dG (\frac {dx}{dG})^2 =-\hf \int dt \frac{1}{\dot G}(\frac {dx}{dt_E})^2\,.
\]

{\bf ii)}

\[
\int d\tau_E  \hf [({\p S_{0}\over \p x})^2 - \hf  \frac{\p^2S_0}{\p x^2}]=-\int dt_E \dot G\hf [({\p S_{0}\over \p x})^2 - \hf  \frac{\p^2S_0}{\p x^2}]\,.
\]

{\bf iii)}
\[
-\int d\tau_E~\frac{ dx}{d\tau_E} {\p S_{0}\over \p x}=\int dt_E~\frac{ dx}{dt_E} {\p S_{0}\over \p x}\,.
\]
So
\be	\label{newrg}
S_E=\int_{t_i}^{t_f} dt~[-\hf \frac{1}{\dot G}(\frac {dx}{dt})^2- \underbrace{ \dot G\hf [({\p S_{0}\over \p x})^2 - \hf  \frac{\p^2S_0}{\p x^2}}_{\frac{\p S_0(x,t)}{\p t}}]-\frac{ dx(t)}{dt} {\p S_{0}(x(t)\over \p x(t)}]\,.
\ee

Note that in writing $\frac{\p S_0(x,t)}{\p t}$, it is understood that the explicit $t$ dependence of $S_0$ due to RG evolution is being differentiated. 
The implicit $t$ dependence due to $x(t)$ is not considered here. The next term 
$\frac{ dx}{dt} {\p S_{0}\over \p x}$ differentiates this implicit $t$-dependence.  Adding them we get $\frac{dS_0}{dt}$.
So we have
\be	\label{newrg2}
S_E=\int_{t_i}^{t_f} dt~\left[-\hf \frac{1}{\dot G}\left(\frac
    {dx}{dt}\right)^2\right] 
 - \left\lbrace S_0(x(t_f),t_f)-S_0(x(t_i),t_i)\right\rbrace
 \ee
Thus, we get for the evolution of $\psi''$:
\[
e^{-S_1 \left(x(t_f),t_f\right)}=\int {\cal D}x\, e^{\hf \int_{t_i}^{t_f}
  dt~\hf \frac{1}{\dot G}(\frac {dx}{dt})^2}e^{-S_0\left(x(t_i),t_i\right)}
e^{-S_1\left(x(t_i),t_i\right)}\,. 
\]
Hence, we obtain
\be  
e^{-S_0\left(x(t_f),t_f\right)-S_1\left(x(t_f),t_f\right)}=\int {\cal
  D}x\, e^{\hf \int_{t_i}^{t_f} dt~\hf \frac{1}{\dot G}(\frac
  {dx}{dt})^2}e^{-S_0\left(x(t_i),t_i\right)-S_1\left(x(t_i),t_i\right)} \,.
\ee
This is of course the usual RG evolution of $\psi'$! Thus in writing \eqref{newrg} what really has been done is that
the evolution of $S_0$ has been introduced into the functional
integral by writing
\begin{align} 
\Delta S_0 &= S_0\left(x(t_f),t_f\right)-S_0\left(x(t_i),t_i\right) =
             \int_{t_i}^{t_f} dt~\frac{dS_0(x(t),t)}{dt}\notag\\ 
& =\int _{t_i}^{t_f} dt~\left[\frac{\p S_0(x(t),t)}{\p t}+\frac{
  dx(t)}{dt} {\p S_{0}(x(t))\over \p x(t)}\right]\,. 
\end{align}

It is important to  point out that \eqref{newrg} is more general than
\eqref{newrg2} because it does not require $S_0$ to satisfy the ERG equation.

Thus, in conclusion, the action governing the evolution of $\psi''$ is
\eqref{newrg}.

\section{Correlation functions in double
  brackets}\label{appendix-double}

In (\ref{double-bracket}) we have defined correlation functions in
double brackets as
\begin{align}
&\vvev{\phi (p_1) \cdots \phi (p_n)} \notag\\
&\equiv \prod_{i=1}^n
\frac{1}{K_\Lambda (p_i)}\cdot \vev{\exp \left( - \frac{1}{2} \int_p
    \frac{K_\Lambda (p)\left(1-K_\Lambda (p)\right)}{p^2}
    \frac{\delta^2}{\delta \phi (p) \delta \phi (-p)} \right)\phi
  (p_1) \cdots \phi (p_n)}_{\SL}\,.
\end{align}
Intuition for this definition partly comes from perturbation theory.
Let us imagine we split the total Wilson action into two parts:
\begin{equation}
S_\Lambda [\phi] = S_{F,\Lambda} [\phi]+ S_{I,\Lambda} [\phi]\,,
\end{equation}
where
\begin{equation}
S_{F,\Lambda} [\phi] =  - \frac{1}{2} \int_p \frac{p^2}{K_\Lambda (p)} \phi
(p) \phi (-p) 
\end{equation}
is the action for the free massless theory.  The second term contains
the rest of the action.  The massless free propagator given by
$S_{F,\Lambda}$ is
\begin{equation}
\vev{\phi (p) \phi(q)}_{S_{F,\Lambda}} = \frac{K_\Lambda (p)}{p^2}
\delta (p+q)\,.
\end{equation}

Consider a connected part of the correlation functions higher than
two-point.  We find
\begin{equation}
\vev{\phi (p_1) \cdots \phi (p_n)}^{\mathrm{connected}}_{\SL} 
= \prod_{i=1}^n K_\Lambda (p_i)\cdot
\left(\textrm{$\Lambda$-independent}\right)\,.
\end{equation}
Each factor of a cutoff function is due to the external propagator.
The $\Lambda$-dependence of the internal propagators is canceled by
the $\Lambda$-dependence of $S_{I,\Lambda}$: this is how the ERG
differential equation (\ref{ERGdiffeq}) is derived.  Hence,
\begin{equation}
\vvev{\phi (p_1) \cdots \phi (p_n)}^{\mathrm{connected}}
\equiv \prod_{i=1}^n \frac{1}{K_\Lambda (p_i)}\, \cdot \vev{\phi (p_1)
  \cdots \phi (p_n)}^{\mathrm{connected}}_{\SL}  
\end{equation}
is independent of $\Lambda$. 

The $\Lambda$-dependence of the two-point function is a little tricky:
\begin{equation}
\vev{\phi (p) \phi (q)}_{\SL} = \frac{K_\Lambda (p)}{p^2} \delta (p+q)
+ K_\Lambda (p) \left(\textrm{$\Lambda$-independent}\right) K_\Lambda
(q)\,.
\end{equation}
The multiplication by $1/(K_\Lambda (p) K_\Lambda (q))$ does not work
due to the tree level propagator.  We first make a subtraction at high
momentum:
\begin{equation}
\vev{\phi (p) \phi (q)}_{\SL} - \frac{K_\Lambda (p) \left(1 -
    K_\Lambda (p)\right)}{p^2} \delta (p+q) = K_\Lambda (p)
\left( \frac{1}{p^2} \delta (p+q) +
  \textrm{$\Lambda$-independent}\right) K_\Lambda (q)\,.
\end{equation}
We then obtain
\begin{align}
\vvev{\phi (p) \phi (q)} &\equiv \frac{1}{K_\Lambda (p) K_\Lambda (q)}
\left( \vev{\phi (p) \phi (q)}_{\SL} - \frac{K_\Lambda (p) \left(1 -
    K_\Lambda (p)\right)}{p^2} \delta (p+q) \right)\\
&= \frac{1}{p^2} \delta (p+q) +
  \textrm{$\Lambda$-independent}\,,\notag
\end{align}
which is independent of $\Lambda$.

\section{Derivation of (\ref{Wbar}) from (\ref{Wbar-scale}) or
  (\ref{Wbar-conf})} \label{appendix-Wbar}

Assuming that $\bar{W}$ is quadratic in $\bar{J}$, we can derive
(\ref{Wbar}) either from scale invariance (\ref{Wbar-scale}) or from
conformal invariance (\ref{Wbar-conf}).  For simplicity, we omit the
bars from $\bar{W}$, $\bar{J}$, and dimensionless momenta in the
following.

Let us consider a quadratic functional (action)
\begin{equation}
W [J] = \frac{1}{2} \int_p J(p) J(-p) w (p)\,,\label{quadratic-Wbar}
\end{equation}
where $w(p)$ is a function of $p^2$.  Ignoring the terms independent
of $J$, (\ref{Wbar-scale}) and (\ref{Wbar-conf}) reduce to
\begin{subequations}
\begin{align}
&\int_p \left[ J(-p) \left( - p \cdot \partial_p + \frac{-D-2+\eta}{2}
  \right) \frac{\delta W}{\delta J(-p)} \right.\nn\\
&\left.\qquad + \frac{1}{2} (- p
  \cdot \partial_p + 2 -\eta ) R(p) \cdot \frac{\delta W}{\delta
    J(-p)} \frac{\delta W}{\delta J(p)} \right] =
0\,,\label{appendix-Wbar-scale} 
\end{align}
and
\begin{align}
& \int_p \left[ J(-p) \left( - p_\nu \frac{\partial^2}{\partial
      p_\mu \partial p_\nu} + \frac{1}{2} p_\mu
    \frac{\partial^2}{\partial p_\nu \partial p_\nu} +
    \frac{-D-2+\eta}{2} \frac{\partial}{\partial p_\mu} \right) \left(
    w (p) J(p) \right)\right.\nn\\
&\qquad\left. + \frac{1}{2} \int_p \left(- p \cdot \partial_p + 2 - \eta
\right) R(p) \cdot \frac{\partial}{\partial p_\mu} \frac{\delta
  W}{\delta J(-p)} \cdot \frac{\delta W}{\delta J(p)}\right] = 0\,,
\label{appendix-Wbar-conf}
\end{align}
\end{subequations}
respectively.

\subsection{scale invariance}

Substituting (\ref{quadratic-Wbar}) into (\ref{appendix-Wbar-scale}),
we obtain
\begin{subequations}
\begin{equation}
\int_p J(-p) \left( - p \cdot \partial_p + \frac{-D-2+\eta}{2}
  \right) \frac{\delta W}{\delta J(-p)} = \frac{1}{2} \int_p J(p)
  J(-p) \left(- p \cdot \partial_p -2 + \eta \right) w(p)\,,
\end{equation}
and
\begin{equation}
\frac{1}{2} \int_p (- p
  \cdot \partial_p + 2 -\eta ) R(p) \cdot \frac{\delta W}{\delta
    J(-p)} \frac{\delta W}{\delta J(p)}
= \frac{1}{2} \int_p \left( - p \cdot \partial_p + 2 - \eta \right) R(p)
\cdot w(p)^2 J(p) J(-p)\,.
\end{equation}
\end{subequations}
Hence, we obtain
\begin{equation}
\left( p \cdot \partial_p - 2 + \eta \right) \left(\frac{1}{w(p)} -
  R(p) \right) = 0\,.\label{equation-w}
\end{equation}
This gives
\begin{equation}
w(p) = \frac{1}{\mathrm{const}\, p^{2-\eta} + R(p)}\,.
\end{equation}
(\ref{Wbar}) corresponds to a particular choice of the constant.

\subsection{conformal invariance}

Substituting (\ref{quadratic-Wbar}) into (\ref{appendix-Wbar-conf}),
we obtain
\begin{subequations}
\begin{align}
&\int_p J(-p) \left( - p_\nu \frac{\partial^2}{\partial
      p_\mu \partial p_\nu} + \frac{1}{2} p_\mu
    \frac{\partial^2}{\partial p_\nu \partial p_\nu} +
    \frac{-D-2+\eta}{2} \frac{\partial}{\partial p_\mu} \right) \left(
    w (p) J(p) \right) \nn\\
&\quad = \frac{1}{2} \int_p J(-p) \frac{\partial
    J(p)}{\partial p_\mu} \left( - p \cdot \partial_p - 2 + \eta
  \right) w(p)\,,
\end{align}
and
\begin{align}
&\frac{1}{2} \int_p \left(- p \cdot \partial_p + 2 - \eta
\right) R(p) \cdot \frac{\partial}{\partial p_\mu} \frac{\delta
  W}{\delta J(-p)} \cdot \frac{\delta W}{\delta J(p)}\nn\\
&\quad = - \frac{1}{2} \int_p J(-p) \frac{\partial J(p)}{\partial p_\mu}
\left( p \cdot \partial_p - 2 + \eta \right) R(p) \cdot w(p)^2\,.
\end{align}
\end{subequations}
Hence, we obtain (\ref{equation-w}) again.

\bibliography{paper-v2}

\end{document}